\documentstyle[prb,aps]{revtex}  
\input psfig
\begin{document}  
\newcommand{\beq}{\begin{equation}}    
\newcommand{\eeq}{\end{equation}}      

\title{ \bf Low frequency shot noise in double-barrier resonant-tunneling  
${\text GaAs   /   Al_{x}Ga_{1-x}As}$ structures in a strong magnetic field}  

\author{  
{\O}. Lund B{\o}$^{(1)}$ and Yu. Galperin$^{(1,2)}$}  
\address{  
$^{(1)}$Department of Physics, University of Oslo,  P. O. Box 1048  
Blindern, N 0316 Oslo, Norway,}  
\address{$^{(2)}$ A. F. Ioffe  
Physico-Technical  Institute, 194021 St. Petersburg, Russia,}  
\date{\today}  
\maketitle
\begin{abstract}  
Low frequency shot noise and dc current profiles  for a  
double-barrier resonant-tunneling structure (DBRTS) under a  
strong magnetic field  
applied perpendicular to the interfaces have been studied.  
Both the structures with 3D and 2D emitter have been considered.  
The calculations, carried out with the  
Keldysh  Green's function technique,  show strong dependencies of both  
the current and noise profiles on the bias voltage and magnetic field.  
The noise spectrum appears sensitive to charge accumulation due to
barriere capacitances and both noise and dc-current are
extremely sensitive to the Landau levels' broadening  
in the emitter electrode and  can  be  used  as  a  powerful  tool  to  
investigate  the  latter.  As  an  example, two specific shapes of the  
levels' broadening have been  considered  -  a  semi-elliptic  profile  
resulting  from self-consistent Born approximation, and a Gaussian one  
resulting from the lowest order cumulant expansion.  
\end{abstract}      
  
\pacs{ 73.40.Kp, 73.50.Td }

\section{Introduction}  
  
In recent years, there has been a great interest in resonant  
tunneling  
through double-barrier resonant tunneling structures (DBRTS) (Fig.~1).  
Such structures  
have been in focus of many experimental and theoretical  
investigations  
since its conception by Tsu and Esaki \cite{esaki} and first  
realization of  
negative differential resistance by Sollner {\em et al.}\cite{sollner}. Many  
important characteristics of DBRTS have been intensely studied,  
e.g. dc-properties , phonon assisted tunneling, time dependent processes and  
frequency response.  
Noise properties of  
DBRTS have also been studied both experimentally\cite{li} and  
theoretically\cite{chen3,butt1,davies,runge,levy}.  
At low temperatures and in the presence of transport current,  
shot noise is the dominant source of electrical  
noise. This kind of noise is due to  
 discreteness of the electron charge, and it is sensitive to the  
degree of correlation between tunneling processes. In general, a  
correlation  leads to an additional frequency dependence of shot noise,  
as well as to its suppression below the so-called  full noise, $S(0)=2e|  
I_{\text{dc}}|$ (at $T=0$)\cite{ziel}.  
Here  $S(\omega)$  is  the  noise  spectrum  (see the exact definition  
below), while  $I_{\text{dc}}$  is  the  average  dc  current.   In  a  
mesoscopic  conductor  having  several independent modes of transverse  
motion (channels), the noise is determined by the partial transmission  
probabilities   $T_m$   as\cite{khlus,leso,butt2}  
$S\propto   \sum_m   T_m(1-T_m)$,   while   the  conductance  goes  as  
$G\propto\sum_m  T_m$.  Suppression of the shot noise is thus expected  
in a phase coherent system when the tunneling probabilities are of the  
order unity for open quantum channels.  
  
Our concern is a DBRTS in a strong magnetic field perpendicular to the  
interfaces.   Magnetic   field   is   an  important  tool  for  sample  
characterization because it leads to the formation of  Landau  levels,  
as  well  as  to  drastic modification of electron wave functions. We  
study the situation  when the magnetic field ${\bf B}$ is  applied  
parallel   to  the  tunneling  current  ${\bf  I}$,  as  schematically  
illustrated  in  Fig.~1. In such a configuration, the  magnetic  field  
leads   to   an   effectively   one-dimensional   tunneling   problem.  
Consequently,  both  the  dc  current  and  the noise appear extremely  
sensitive to the details of the density-of-states behavior. We believe  
that such a sensitivity can provide a  powerful  tool  to  investigate  
details  of  the  Landau  levels'  broadening  in  resonant  tunneling  
structures.  
  
The paper is organized as follows: Section~\ref{sec2} describes the  
model Hamiltonian as well as the basic expression from which the current and  
shot noise profiles will be derived in Section~\ref{sec3}. In  
Appendix~\ref{appa} and  Appendix~\ref{appb} the Green's functions  
used in our calculations are expanded.  
  
As an example, we consider a GaAs$^+$/ Al$_{0.3}$Ga$_{0.7}$As/ 
GaAs/ Al$_{0.3}$Ga$_{0.7}$As/ GaAs$^+$  
DBRTS, with the barriers' and the well widths of the  order  of  40-60  
{\AA}. Such structures were extensively studied experimentally.  
In  many  cases the barrier height is about 300 meV, and it is  
assumed that there exists only one quasi-bound state in the well.  
  
\section{Model and  basic expressions}  
\label{sec2}  
  
Consider a DBRTS in the presence of an external magnetic  field  ${\bf  
B}$  perpendicular  to the interfaces which are assumed perfect, ${\bf  
B} \parallel {\bf I}\parallel {\bf z}$. Within the quantum  well,  the  
electron  wave function can be expressed as a product of a quasi-bound  
state $\chi(z)$ times a wave function correspondent to the  motion  in  
the   $x-y$  plane.  Let  us  denote  the  energy  of  the  motion  in  
$z$-direction  as  $\epsilon_0$.  Under   the   Landau   gauge   ${\bf  
A}=(0,Bx,0)$  the  wave  functions can be specified by  
the set of quantum numbers $\alpha=(n,k_y)$ as  
\beq \phi_{\alpha}({\bf r}) = {1\over \sqrt{Ly}}  
\exp(ik_y y)   \varphi_n (x + l^2 k_y)   \chi(z). \eeq  
The  corresponding  energy  levels  (measured from the conduction band  
edge) are  
\begin{equation} E_{\alpha}= E_n = \epsilon_{0} +  
\hbar \omega_c (n + \frac{1}{2}). \end{equation}  
Here, $\varphi_n(x)$ denote harmonic oscillator states,  
$\omega_c \equiv eB/m^*$ is  
the cyclotron frequency, and $l \equiv \sqrt{\hbar/eB}$ is the  Landau  
magnetic length.  
  
Similarly,  electron  states in the leads are specified by the quantum  
numbers $\beta =(m,k_{j,y} , k_{j,z})$, where $j\equiv e(c)$ refers  to  
emitter(collector)   states,   respectively.  The  corresponding  wave  
functions and energy levels under the bias $eV$ are given as:  
\begin{equation}  
\phi_{j,\beta}({\bf r})={1\over\sqrt{L_y  L_z}}  
\exp (ik_{j,z} z +ik_{j,y} y)\varphi_m(x+l^2 k_{j,y}), \end{equation}  
\begin{equation} E_{j,\beta} =  
\frac{(\hbar k_{j,z})^2}{2m^{*}} +\hbar \omega_c (m + \frac{1}{2})  
+a_jeV  
\end{equation}  
where  $0<a_e<1$ and $a_c=a_e-1$ (the symmetric $a_e=0.5$ case will be  
considered in our numerical calculations).  We  arrive  at  the  model  
Hamiltonian  
\beq  
{\cal     H}     ={\cal     H}_{\text     e}+{\cal     H}_{\text    T}  
=\sum_{j,\beta}E_{j,\beta}        c_{j,\beta}^\dagger      c_{j,\beta}  
+\sum_\alpha  E_\alpha  c_\alpha^\dagger  c_\alpha  +  \sum_{j,\alpha,  
\beta}\left    [V_{j,\beta\alpha}     c_\alpha^\dagger     c_{j,\beta}  
+h.c\right],  
\eeq  
where  the  tunnel  matrix  elements  $V_{j,\beta\alpha}$  have  to be  
calculated using the eigenstates listed above.  Since  the  interfaces  
are  assumed  to  be perfect, the quantum numbers $n$ and $k_y$ are  
conserved during the tunneling process, and so the calculation of  the  
matrix  elements  $V_{j,\beta\alpha}$  reduces  to  the  solving  of a  
one-dimensional Schr\"{o}dinger equation\cite{zrc}, following with the  
application of the Bardeen's prescription\cite{bar}. Consequently, the  
tunneling matrix elements can be written as  
\beq\label{vmatr}  
V_{j,\beta\alpha}=\delta_{m,n}  \delta(k_y-k_{j,y})  
V_{j,n}(k_{j,y},k_{j,z}). \eeq  
In noise calculations the time dependence of  the  tunneling  currents  
flowing  through  the  DBRTS  is  important,  and  hence  the junction  
capacitances should be taken into account. The effect of the  junction  
capacitances  can  be  included  in  our  model  with  the  help of an  
equivalent circuit of the DBRTS as shown in Fig.~2.\cite{ingo}  
In this circuit,  
we specify the  currents  through  the  emitter  (collector)  barriers  
$I_{\text {e(c)}}(t)$ and their resistances as $R_{\text {e(c)}}$.  The  
``external'' current $I(t)$ is in this model given by  
\beq I(t) =\label{itot}  
{C_{\text c}\over C}I_{\text e}(t)+{C_{\text e}\over C}I_{\text c}(t),  
\eeq  
where $C_{\text {e(c)}}$ is the  
capacitance of the emitter (collector) barrier  and $C=C_{\text e}+  
C_{\text c}$ is the total capacitance of the quantum well.  
In the symmetric case $C_{\text e}=C_{\text c}$, we arrive at  
the simple  
relation $I(t)=[I_{\text e}(t)+I_{\text c}(t)]/2$, which was the basis  
of the Chen~\&~Ting's\cite{chen3} calculation for shot noise in  
a DBRTS in a zero magnetic field.  
If  one  ignores  the  charge  accumulation, all three currents are the  
same\cite{butt1},  
$I(t)=I_{\text e}(t)=I_{\text c}(t)$ and  
the result in this case can be obtained from the following formulas  
in the limit of strong asymmetry, $C_{\text {e(c)}}/C \rightarrow 0$.  
The asymmetry in capacitances is  of  course  not  important  for  the  
dc-current, where  
\beq \label{idccons}  
I_{\text dc}=I_{\text dc,e}=I_{\text dc,c}\eeq  
In the further analysis it is convenient  
to out $\hbar = 1$,  
and   then   restore  $\hbar$  again  in  the  final  expressions  and  
order-of-magnitude estimates.  
  
The tunneling current $I_{\text e}$ flowing into the well from the emitter  
and the current $I_{\text c}$ flowing out of the well to the collector, are  
in general different. They are  
 given by the expressions  
\begin{eqnarray}\label{icom}I_{\text dc,j}&=& -e  \kappa_j\langle  
\dot{N}_j(t)\rangle =  
-ie  \kappa_j\left\langle[{\cal H}_{T}(t),N_{\text j}(t)]\right\rangle  
\nonumber\\  
&=& -i  2e  \kappa_j\sum_{\beta, \alpha}\left[V_{j,\beta\alpha}  
\langle c_\alpha^\dagger (t)   c_{j,\beta}(t)\rangle-V_{j,\beta\alpha}^*  
\langle c_{j,\beta}^\dagger(t)   c_\alpha (t)\rangle \right]  
\end{eqnarray}  
where $N_j(t)$ are the Heisenberg number-of-particles operators,  
$\kappa_e\equiv 1$, $\kappa_c\equiv -1$, and  
a spin degeneracy factor  $2$ is introduced.  
  
The shot noise spectrum is defined as the Fourier transform of the  
current-current auto-correlation function as\cite{ziel}  
\beq\label{four} S(\omega )=2\int_{-\infty}^\infty S(t) e^{i\omega t}dt  
=4\int_0^\infty S(t) \cos(\omega t)  dt\eeq  
where $S(t)$ is the quantum mechanical and statistical average of  
the current-current anti-commutator:  
\beq \label{s1}S(t)= {1\over 2}\left\langle\{\Delta I(t)\, ,\,\Delta  
I(0)\}\right\rangle = {1\over 2}\left\langle\{I(t)\, ,\, I(0)\}\right  
\rangle - I_{\text{dc}}^2.\eeq  
{}From (\ref{itot}) and  (\ref{icom}), it  
can be expressed  (having in mind the spin degeneracy factor of 2) as  
\begin{eqnarray}  
\label{s2}  
S(t)=-e^2\sum_{j,j_0,\alpha , \alpha_0 ,\beta ,\beta_0}  
\eta_j   \eta_{j_0}&\times& \left[V_{j,\beta\alpha}V_{j_0,\beta_0\alpha_0}  
\left\langle\left\{c_\alpha^\dagger (t)c_{j,\beta}(t),  
c_{\alpha_0}^\dagger (0)c_{j_0,\beta_0}(0)\right\}\right\rangle\right.  
\nonumber\\  
&-&V_{j,\beta\alpha}V_{j_0,\beta_0\alpha_0}^*  
\left\langle\left\{c_\alpha^\dagger (t)c_{j,\beta}(t),  
c_{j_0,\beta_0}^\dagger (0) c_{\alpha_0} (0)\right\}\right\rangle  
\nonumber\\  
&-& V_{j,\beta\alpha}^*V_{j_0,\beta_0\alpha_0}  
\left\langle\left\{c_{j,\beta}^\dagger (t) c_\alpha (t),  
c_{\alpha_0}^\dagger (0)c_{j_0,\beta_0}(0)\right\}\right\rangle  
\nonumber\\  
&+& \left. V_{j,\beta\alpha}^*V_{j_0,\beta_0\alpha_0}^*  
\left\langle\left\{c_{j,\beta}^\dagger (t) c_\alpha (t),  
c_{j_0,\beta_0}^\dagger (0) c_{\alpha_0}  
(0)\right\}\right\rangle\right].  
\end{eqnarray} where $\eta_e\equiv C_c/C$ and  $\eta_c\equiv  -C_e/C$.  
Being expressed through Feynman's graphs, these  
averages  involve only the diagrams with the Green's functions  
connecting the times $t$ and $0$,  
since disconnected parts are all canceled by the  
$I_{\text{dc}}^2$ subtraction in (\ref{s1}).  
  
\section{The results}  
\label{sec3}  
  
The task is now to expand the quantum statistical averages appearing in  
(\ref{icom}) and (\ref{s2}).  
For a finite bias, the DBRTS as a whole is not in thermal equilibrium, and it  
seems thus appropriate to employ the Keldysh non-equilibrium Green's function  
technique\cite{lifs,mahan}, where the two lead subsystems  are supposed to  
have their own local equilibrium.  
  
Expanding (\ref{icom}) yields (Appendix~\ref{appa})  
\beq \label{idc3} I_{\text dc,j}= -{e\;g_B\over \pi}  
\sum_{n}\int_{-\infty}^{\infty}d\varepsilon\;  
\gamma_j(n,\varepsilon)   A(n,\varepsilon)  
\left[f_{QW}(\varepsilon)-f_j(\varepsilon)\right].\eeq  
In the above expression, $g_b\equiv L_x\:L_y / 2\pi l^2$ is the magnetic
$k_y$ summation degeneracy factor, 
$\gamma_j(n,\varepsilon)$ is the escape rate  
to the lead $j$,  
$f_{QW}(\varepsilon)$  and  $f_j(\varepsilon)$  are   the   occupation  
factors, while  
$A(n,\varepsilon)$ is the spectral function for  $n$th  Landau  
level in the well,  
\beq  
A(n,\varepsilon)=-2  \Im    G_R(n,\varepsilon)=  
{\gamma(n,\varepsilon )\over (\varepsilon - E_n)^2 + [\gamma(n,\varepsilon)/2]^2}.  
\eeq  
Here, $G_R(n,\varepsilon)$ is the retarded electron Green's function, and  
$\gamma(n,\varepsilon)=\gamma_e (n,\varepsilon)+ \gamma_c(n,\varepsilon)$  
is the level broadening due to the finite escape rate to the leads.  
Usually,  the  energy  distance between the resonant level in the well  
and the tops of the barriers is much greater than the escape rate  
from the well, $\gamma$. In this case the tunneling matrix element  
(\ref{vmatr})  
can be considered as a smooth function of the energy in comparison  with  
the energy dependence of the density-of-states in the leads,  
$$g_j(n,\varepsilon) \equiv \sum_{k_{j,z}}  
\delta(\varepsilon - E_{j,\beta}).$$  
Thus the escape rates $\gamma_j$ can be expressed as  
$$\gamma_j (n,\varepsilon) = 2\pi |V_j|^2g_j(n,\varepsilon).$$  
Consequently,   the   noise   appears   a   sensitive  tool  to  study  
density-of-states in the electrodes in a  magnetic  field.  Below,  we  
will  do  numerical  calculations  for  two  models for the density of  
states - for a constant Lorentzian broadening, and for the  so-called  
self-consistent Born approximation.  
  
Since both leads are assumed to be in a  
thermal equilibrium with different electro-chemical potentials and  
Fermi energies, the  
occupation numbers can be expressed as the Fermi functions:  
\beq f_j(\varepsilon)={1\over{e^{(\varepsilon-E_f-a_jeV)/k_B T}+1}}.\eeq  
However,  thermal  equilibrium is not maintained in the quantum well and  
thus one cannot use the Fermi distribution for the electrons in this region.  
Instead, from the dc-current conservation law  
(\ref{idccons}), the  
weighed average occupation factor is determined as\cite{jons}  
\beq \label{fqw}  
f_{QW}(n,\varepsilon)={\gamma_e (n,\varepsilon)   f_e (\varepsilon)  
+ \gamma_c (n,\varepsilon)   f_c (\varepsilon)  
\over \gamma(n,\varepsilon)}.\eeq  
Re-introducing $\hbar$ to return to the proper units, we arrive  
at the Landauer formula \cite{land}  
\beq \label{iland}I_{\text dc}= {e\; g_B\over \pi\hbar}\sum_n\int d\varepsilon\;  
T_n(\varepsilon)\left[ f_e(\varepsilon)-f_c(\varepsilon)\right]\eeq  
with the transmission probability  
\beq \label{trans}  
T_n(\varepsilon)\equiv {\gamma_e(n,\varepsilon)  \gamma_c(n,\varepsilon)\over  
\gamma(n,\varepsilon)}  
A(n,\varepsilon)\eeq  
  
To get a relatively simple expressions for the shot  noise from 
Eq.~(\ref{s2})  
we assume the following approximations.  First,  we  assume  that  the  
resonant   level   is   situated  well  inside  the  resonant tunneling  
region,  
\begin{eqnarray} \label{in1}  
|a_j  eV_j  +E_F  -  E_n|  \gg     \max( \hbar \omega,\gamma  
,\nu_j),  
\nonumber\\  
|a_j eV_j - \epsilon_0|  \gg     \max(  \hbar  \omega,\gamma  
,\nu_j),  
\end{eqnarray}  
and  that $\omega \ll \omega_c$. These inequalities allow us  
to put $f_j(\varepsilon\pm\omega)\rightarrow  
f_j(\varepsilon)$ and $\gamma_j(n,\varepsilon\pm\omega)\rightarrow  
\gamma_j(n,\varepsilon)$.  
Second, the temperature is assumed to be low ($k_BT\ll\gamma$), in which case  
the Fermi functions can be approximated as step functions.  
Keeping those approximations in mind, we arrive at the following  
result (Appendix~\ref{appb}):  
\begin{eqnarray}\label{s4}  
S(\omega)=S(-\omega)&=& {e^2\; g_B\over\pi}\sum_n\int d\varepsilon  
\left[f_e(\varepsilon)-f_c(\varepsilon)\right]^2  \times\nonumber\\  
&\times&\left\{ A(n,\varepsilon)A(n,\varepsilon-\omega)  
\left[  
{\gamma_e  \gamma_c  (\gamma_e C_e - \gamma_c C_c)  
(\gamma_e C_c - \gamma_c C_e)\over  
C^2  \gamma^2}- {\gamma_e^2\gamma_c^2\over \gamma^2}\right]\right.\nonumber\\  
&+& \left[A(n,\varepsilon)+A(n,\varepsilon-\omega)\right]  
{C_e^2+C_c^2\over C^2}  
{\gamma_e\gamma_c\over\gamma}\nonumber\\  
&-&\left. 4   \Re\left[ G_R(n,\varepsilon)\right]  
\Re\left[G_R(n,\varepsilon-\omega)\right]  
{C_e  C_c \over C^2}  \gamma_e\gamma_c\right\}  
\end{eqnarray}  
where $\gamma_j\equiv \gamma_j(n,\varepsilon)$.  
  
Re-inserting $\hbar$, and using the relation  
$4\,\gamma_e \gamma_c \Re\left[ G_R(n,\varepsilon)\right]^2=  
4\, T_n(\varepsilon)-(\gamma^2/\gamma_{\text e}\gamma_{\text c})  
\; T_n^2(\varepsilon)$, we arrive at the well known result:  
\beq\label{s0}  
S(0)={2e^2\; g_B\over\pi\hbar}\sum_n\int d\varepsilon\;  
T_n(\varepsilon)\left[1-T_n(\varepsilon)\right]  
\left[f_e(\varepsilon)-f_c(\varepsilon)\right]^2
\eeq 
As one could expect, the zero frequency shot noise does thus not depend
on the barriere capacitances and the above result coincides with previous
calculations which have been performed for point contacts \cite{khlus,leso}, 
for arbitrary phase coherent
two terminal conductors\cite{butt2} 
(neglecting barriere capacitances), and also 
for a DBRTS in the regime  of  incoherent  
tunneling\cite{davies}. The main features of our problem  is  that  the  
combinations  $T_n(1-T_n)$  enter  for each Landau level independently  
and that the tunneling probabilities $T_n$  are  strong  functions  of  
magnetic field.  An important  feature  is  that  
Eq.~(\ref{s0}) holds even if the inequality (\ref{in1})  is  violated.  
That  makes  zero-frequency  shot noise, together with the dc-current,
a powerful tool to investigate  
the density of states in the leads which manifests itself through  the  
escape rates $\gamma_j$.  
  
The results for a particular DBRTS device are shown in Fig.~3.  
Here  we use the model of constant Lorentzian broadening of the Landau  
levels, where  the escape rates can be expressed as  
(Appendix~\ref{appa})  
\beq\label{escape}  
\gamma_j(n,\varepsilon)= {\Upsilon_j \; \nu\over 2\sqrt{2}\left\{\left[\left(  
\varepsilon-E_{j,n}\right)^2 +\left(\nu /2\right)^2\right]  \left[  
\sqrt{\left(\varepsilon-E_{j,n}\right)^2+\left(\nu /2\right)^2}  
+E_{j,n}-\varepsilon\right]\right\}^{1/2}}.  
\eeq  
Here, $\Upsilon_j$ is a constant characterizing the  strength of the  
escape rate and $E_{j,n}\equiv eVa_j+\omega_c(n+1/2)$.  
Note that there are   peaks  in   the   dimensionless   shot   noise  
factor  $S(\omega)/e I_{\text{dc}}$ at the voltages when an intra-well  
Landau  level passes the  emitter's  electro-chemical  potential.  Those  
peak's  shape  is  determined  by  an  interplay  between the  quantum  
suppression ($S(0) \propto T_n(1-T_n)$) and a finite  
broadening of the  Landau levels in the quantum well. In addition,  
a small  peak appears in the  
dc-current curve at the end of the resonant tunneling region (in  
our example, at  $eV \sim 55$ meV) due to  the finite  
broadening of the lead electron states. This broadening  
can typically be of the size $\nu\sim\hbar e /m^*\mu\sim 0.5$ meV  
($\mu$ is the electron mobility).  
  
The effect  of  the  level  broadening  in  the  leads  is  even  more  
pronounced in the case of a 2D-emitter.  
For  numerical  calculations  in  this  case  we  employ the so-called  
self-consistent Born approximation \cite{ando}.  
In this approximation, the density of states takes a  
semi-elliptic form and the escape rate is then given by:  
\beq  
\gamma_e(n,\varepsilon)=\Upsilon_e   {4\;\hbar\over\sqrt{2m^*}   L_{ez}\;   
\nu }  \sqrt{ 1-\left({\varepsilon - E_{e,n}\over\nu}\right)^2 }.\eeq  
where $E_{e,n}=eVa_e+\omega_c(n+1/2)+\epsilon_e$, $\epsilon_e$  
is the emitter quasi-bound level and $L_{ez}$ is the width of the 2D-emitter.
The lead broadening depends in  
this case on the magnetic field and is given  
by $\nu  \sim  \sqrt{2\hbar^2  e\;\omega_c  /  \pi  
m^*\mu}$, where $\mu$ is the mobility of the 2DEG.  
In our example,  
$\mu\sim 10^6$ cm$^2$/Vs, at  
$\hbar\omega_c=10$ meV we get $\nu\sim 0.35$ meV.  
In realistic systems, sharp  
edges  of  the  semi-elliptical density-of-states profile are smoothed,  
the smoothing for a long-range potential being  Gaussian\cite{rai}. To  
check the sensitivity to the smoothing we made also calculations for a  
Gaussian density-of-states profile.
The calculations show that both the  
current and the noise profiles can be very  
sensitive to the degree of such a smoothing.  
Fig.~4 shows the dc-current and zero frequency shot noise results for  
a particular  DBRTS device with 2D emitter calculated according to  
the self-consistent Born approximation (semi-elliptic profile) as well  
as for a Gaussian  profile obtained from a  
so-called lowest-order cumulant approximation\cite{ando,gerh}.  
A double-peak structure is obtained  
with the Gaussian profile in contrast to the single peak appearing  
in the case of a semi-elliptic profile.  
  
We believe that our results can serve as a basis  
for an experimental  test  of  the  strength  of  the  Landau  level's  
smearing by impurities.  
In our example, the splitting of the noise and current peaks in the case  
of the Gaussian level broadening case is about 2 meV, and should  be  
observable at temperatures $T\ll 20$ K.  
  
Finally,  we  give  an  expression  for the shot noise valid at finite  
frequency provided the inequality (\ref{in1}) holds.  
Integrating (\ref{iland}) and (\ref{s4}) with a 3D emitter,  
we arrive at the expression (Appendix~\ref{appb})  
\begin{eqnarray}\label{stot1}  
S(\omega)={2| e I_{dc}|\over C^2}  
\left\{\left(C_e^2+C_c^2\right)+{1\over\gamma^2+\omega^2}\right.
&\times&\left[C_eC_c(\gamma_e^2+\gamma_c^2)
-(C_e^2+C_e^2)\gamma_e\gamma_c\right.\nonumber\\ 
&-&\left.\left.\gamma_e\gamma_c C^2+C_eC_c\gamma^2\right]\right\}  
\end{eqnarray} 
This result is strongly dependent of the bias voltage because  of  the  
voltage dependence of the escape rates.  
Indeed, at $|\epsilon_0 -eVa_j| \gg \max(\hbar\omega,\gamma, \nu_j )$,  
$$  
\gamma_j\equiv\gamma_j(eV)=\Upsilon_j  {\Theta (\epsilon_0 -eVa_j)  
\over \sqrt{\epsilon_0 -eVa_j}}.  
$$  
As our two special cases, symmetric capacitance ($C_e=C_c$) and  
no charge accumulation ($C_{c(e)} \rightarrow 0$), we arrive at the  
relations similar to those obtained by  
Chen~\&~Ting\cite{chen3} and by  B\"{u}ttiker\cite{butt1} in  
zero magnetic field:  
\begin{eqnarray}  
\label{sw1}  
S_{\text sym}(\omega)&=&\mid e I_{dc}\mid  
\left[1+{\gamma^2\over\gamma^2+\omega^2}  
\left(1-4{\gamma_e\gamma_c\over\gamma^2}\right)\right]\\  
\label{sw2}  
S_{\text asym}(\omega)&=&2\mid e I_{dc}\mid  
\left[1-2{\gamma_e\gamma_c\over\gamma^2+\omega^2}\right].  
\end{eqnarray}  
However,  the  important   difference  is    strong dependencies of the  
escape rates on both electric and magnetic fields.  
The frequency dependency of the noise in those two cases are very different  
(Fig.~5) and can serve as a basis for an experimental test  
of the importance of the charge accumulation on the barrier capacitances  
in the DBRTS tunneling structure.  
  
The present work has partially been supported by the Norwegian Research  
Council, Grant No. 100267/410.  
  
\appendix  
\section{Green's function expansion for dc current}  
\label{appa}  
  
The quantum statistical averages appearing in (\ref{icom}) is expanded  
using the  
Keldysh non-equilibrium Green's function technique\cite{lifs,mahan}.  
Four different Green's functions, appropriate  
for a S-matrix expansion in the time-loop formalism,  
are defined along a closed  
time path that runs from $-\infty$ to $+\infty$ along the  
$\sigma =$`1'  
branch and then returns from $+\infty$ back to $-\infty$ along the  
$\sigma =$ `2' branch:  
\beq \label{grdef}G_{\sigma_1\sigma_2}(t_1-t_2)=-i\langle  
{\cal T}_t  c(t_1)c^\dagger(t_2)\rangle\eeq  
where by $\sigma_n=  1 (2)$ is meant that $t_n$  
is located on the `1'(`2') branch and ${\cal T}_t$ is the generalized  
chronological operator ordering physical operators along the closed time  
path. In the Fourier transformed  energy space, the Green's functions  
are simply related to the retarded Green's functions as:  
\begin{eqnarray}\label{grrel}  
G_{11}(\varepsilon)&=&i f(\varepsilon) A(\varepsilon)  
+ G_R(\varepsilon),\nonumber\\  
G_{12}(\varepsilon)&=&i f(\varepsilon) A(\varepsilon),\nonumber\\  
G_{21}(\varepsilon)&=&-i \left[1-f(\varepsilon)\right]  
A(\varepsilon),\nonumber\\  
G_{22}(\varepsilon)&=&-i \left[1-f(\varepsilon)\right]  
A(\varepsilon) - G_R(\varepsilon).  
\end{eqnarray}  
Here $A(\varepsilon)\equiv -2   \Im \left[G_R(\varepsilon)\right]$  
is the spectral function, while  $f(\varepsilon)$  is  the  occupation  
number in the  
region considered. The following retarded quantum well and lead Green's 
functions  are used as the basis in the calculations:  
\begin{eqnarray}\label{green}  
G^0_R(\alpha, \varepsilon)&=&[ \varepsilon-E_\alpha +  
i  \gamma (n,\varepsilon)/2]^{-1}, \nonumber\\  
G^0_R(j\beta, \varepsilon)&=&[ \varepsilon-E_{j,\beta}  
+i  \nu_j /2]^{-1}. \end{eqnarray}  
Here $\gamma(n,\varepsilon)=\gamma_e(n,\varepsilon)+\gamma_c(n,\varepsilon)$  
is the broadening of the resonant states due to the finite tunneling rate  
to  the  leads,  and $\nu_j$ is the broadening of electron states in the  
leads  due to electron scattering.  
  
The dc current is expanded to lowest order in the time-loop S-matrix  
expansion\cite{mahan}, which from (\ref{icom}) yields  
(as diagrammatically represented in Fig.~6):  
\begin{eqnarray}  
I_{\text{dc},j}&=&  
-4e  \kappa_j\lim_{\acute{t}\rightarrow t}\sum_{\alpha,\beta}\int_{-\infty}^  
{-\infty} dt_1\mid V_{j,\beta\alpha}\mid^2\Re\left\langle  
{\cal T}_t c_{j,\beta}^\dagger(t_1)c_\alpha(t_1)  
c_{j,\beta}(t)c_\alpha^\dagger(\acute{t})\right\rangle\nonumber\\  
&=&\quad 4e\sum_{\alpha,\beta}\int_{-\infty}^{\infty}dt_1  
\mid V_{j,\beta\alpha}\mid^2 \;
\Re \left[  G_{11}(j\beta, t-t_1)G_{12}(\alpha,t_1-t)\right.\nonumber\\  
&&\qquad\qquad\qquad\qquad\qquad\quad\left. -G_{12}(j\beta, t-t_1)G_{22}(\alpha,t_1-t)  \right].  
\end{eqnarray}  
In the first of the above integrals, $t$ is located on the `$1$' branch,  
$\acute{t}$ is located on  
the `$2$' branch and the $t_1$ integral is taken along the time-loop from  
$-\infty$ to $+\infty$ and back to $-\infty$. The latter result, introducing  
Green's functions according to (\ref{grdef}),  
is expressed as an integral over ordinary real time axis from $-\infty$  
to $\infty$. The Fourier transform of  this result, with the substitution  
of (\ref{grrel}), yields:  
\beq \label{idc2}  
I_{\text dc,j}= -{e\over \pi}\sum_{\alpha,\beta}\mid V_{j,\beta\alpha}\mid^2  
\int_{-\infty}^{\infty}d\varepsilon\;  
   A(\alpha,\varepsilon)   A(j\beta,\varepsilon)  
\left[f_{QW}(\varepsilon)-f_j(\varepsilon)\right], \eeq  
where  
$f_{QW}(\varepsilon)$ and $f_j(\varepsilon)$ are respectively the  
occupation numbers in the  
quantum well and leads. Using the escape rates from the 
quantum well states to the lead $j$, defined as  
\beq \gamma_j(n,\varepsilon)=\sum_{k_{j,z}}\mid V_j\mid^2  
A(j\beta,\varepsilon)\eeq  
where the tunneling matrix  
elements in (\ref{vmatr}) have been assumed to be 
independent on any quantum numbers  
($V_{j,n}(k_y^j, k_z^j)=V_j$) and taking into account  
the $k_y$ independence of the electron Green's functions,  
($A(\alpha,\varepsilon)=A(n,\varepsilon)$), we arrive at 
(\ref{idc3}) and (\ref{escape}).  
  
\section{Green's function expansion for shot noise}  
\label{appb}  
  
The quantum statistical averages appearing in (\ref{s2}) is expanded in a similar way as with the  
dc current. It is found that  
$S(\omega)$ is symmetric in $\omega$ and  
can be written as a sum of 6 different  
terms (represented by the diagrams in Fig.~7):  
\beq S(\omega)=S(-\omega)=S_1(\omega)+S_2(\omega)+S_3(\omega)  
+S_4(\omega)+S_5(\omega)+S_6(\omega).\eeq  
$S_1(\omega)$ is expanded from the first term in (\ref{s2}) as  
\beq\label{s1a} S_1(\omega)=S_{1a}(\omega)+S_{1a}(-\omega)\eeq  
with  
\begin{eqnarray}\label{s1b}  
S_{1a}(\omega)=&-&{e^2\over \pi}\sum_{j,j_0,\alpha ,\beta}\eta_j  \eta_{j_0}  
\mid V_j\mid^2\mid V_{j_0}\mid^2\int d\varepsilon  \sum_{\sigma_1,\sigma_2}  
(-1)^{\sigma_1+\sigma_2}\nonumber\\  
&\times&\left[G_{\sigma_21}(\alpha,\varepsilon)  
G_{2\sigma_2}(j\beta,\varepsilon)  
G_{\sigma_12}(\alpha,\varepsilon-\omega)  
G_{1\sigma_1}(j_0\beta,\varepsilon-\omega)\right].  
\end{eqnarray}  
$S_2(\omega)$ is the contribution from the from the  4'th term in (\ref{s2}),  
simply related to $S_1(\omega)$ as  
\beq S_2(\omega)=S_{1}^* (\omega).\eeq  
The second term in (\ref{s2}) has both zeroth [$S_3(\omega)$]  
and second order [$S_4(\omega)$] contributions:  
\begin{eqnarray}  
S_{3}(\omega)&=&  
{e^2\over \pi}\sum_{j,\alpha ,\beta}\eta_j^2  \mid V_j\mid^2  
\int d\varepsilon  
G_{21}(j\beta,\varepsilon)  G_{12}(\alpha,\varepsilon -\omega)  
+G_{12}(j\beta,\varepsilon)  G_{21}(\alpha,\varepsilon -\omega)  
\nonumber\\  
S_4(\omega)&=&{e^2\over \pi}\sum_{j,j_0,\alpha ,\beta}\eta_j  \eta_{j_0}  
\mid V_j\mid^2\mid V_{j_0}\mid^2\int d\varepsilon  
\sum_{\sigma_o\sigma_t=\{12,21\}}  \sum_{\sigma_1,\sigma_2}  
(-1)^{\sigma_1+\sigma_2}\times\nonumber\\  
&\times&\left[G_{\sigma_1\sigma_0}(j\beta ,\varepsilon)  
  G_{\sigma_2\sigma_1}(\alpha,\varepsilon)  
G_{\sigma_t\sigma_2}(j_0\beta,\varepsilon)  
G_{\sigma_0\sigma_t}(\alpha,\varepsilon-\omega)\right].  
\end{eqnarray}  
The zeroth [$S_5(\omega)$] and second order [$S_6(\omega)$]  
contributions from the third term in  
(\ref{s2}) are simply given as:  
\begin{eqnarray}  
S_5(\omega)=S_3(-\omega)\nonumber\\  
S_6(\omega)=S_4(-\omega).  
\end{eqnarray}  
The diagrams of the type shown in Fig.~8 are not  taken  into  account  
explicitly because they are  
already included in $S_3(\omega)$ and $S_5(\omega)$, since the quantum well  
electron Green's functions we use as our basis  
are originally dressed by tunneling to the leads\cite{runge}.  
Summing up the different diagrammatic terms we neglect the contribution 
from real part of  
the lead retarded Green's functions. This is a reasonable approximation  
since it corresponds to a Hilbert transform  
of the imaginary part (proportional to the escape rates)  
and it appears  
that, in and above the resonant tunneling region,  
its contribution is negligible. Keeping this in mind, as well as  the  
approximations listed in the main text ($f_j(\varepsilon\pm\omega)\rightarrow  
f_j(\varepsilon)$, $\gamma_j(n,\varepsilon\pm\omega)\rightarrow  
\gamma_j(n,\varepsilon )$ and $k_BT\ll  \gamma$), we arrive  
at the result (\ref{s4}).  
  
Integrating the shot noise expression (\ref{s4}) and the dc-current  
(\ref{iland}), we make use of the following  
integrals over the resonant tunneling region, valid for  
a Landau level located well inside the  
resonant tunneling region according to (\ref{in1}):  
\begin{eqnarray}  
\int d\varepsilon  A(n,\varepsilon)  &\approx& 2\pi\nonumber\\  
\int d\varepsilon  A(n,\varepsilon)  A(n,\varepsilon-\omega)  
&\approx&{4\pi\gamma\over\gamma^2 +\omega^2}\nonumber\\  
\int d\varepsilon  {\text Re}\left[G_R(n,\varepsilon)\right]  
{\text Re}\left[G_R(n,\varepsilon-\omega)\right]  
&\approx& {\pi\gamma\over\gamma^2+\omega^2}.  
\end{eqnarray}  
With those relations, we arrive at (\ref{stot1}).

\begin{figure}\label{fig1} 
\centerline{\psfig{figure=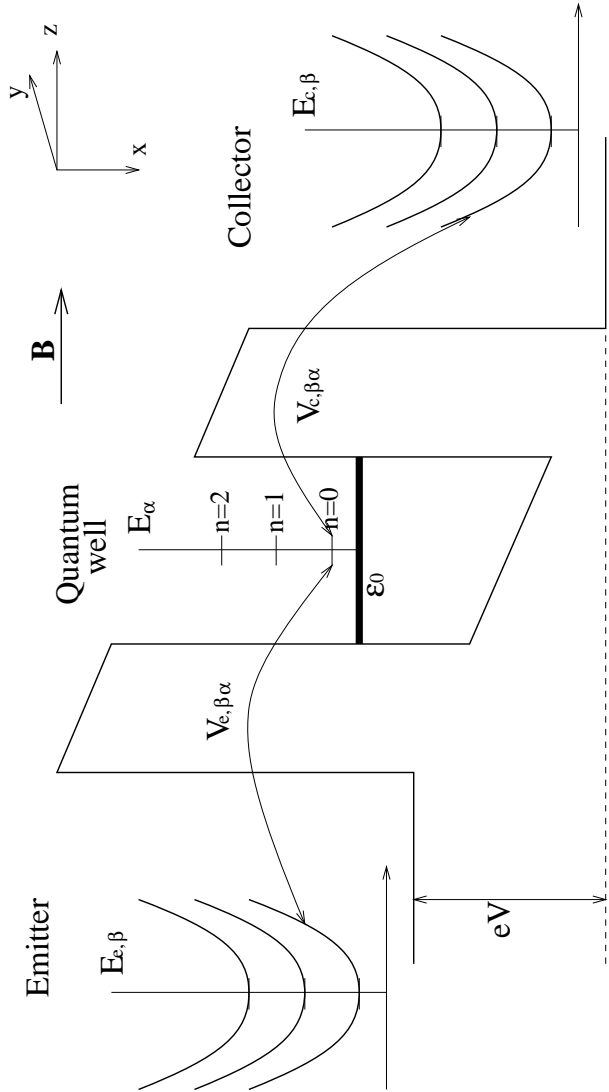,width=8cm}}
\vspace{0.5cm}  
\caption{Schematic illustration of the  
double-barrier resonant tunneling structure (DBRTS).}
\end{figure}  
  
\begin{figure}\label{fig2}
\centerline{\psfig{figure=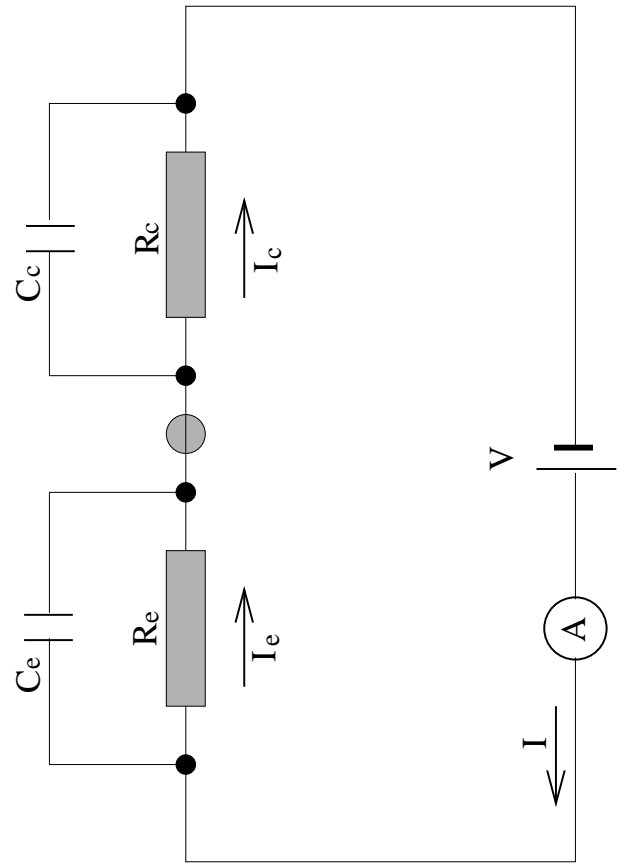,width=8cm}}  
\vspace{0.5cm}  
\caption{Equivalent circuit for a DBRTS.}
\end{figure}  
\newpage
  
\begin{figure}\label{fig3}    
\centerline{\psfig{figure=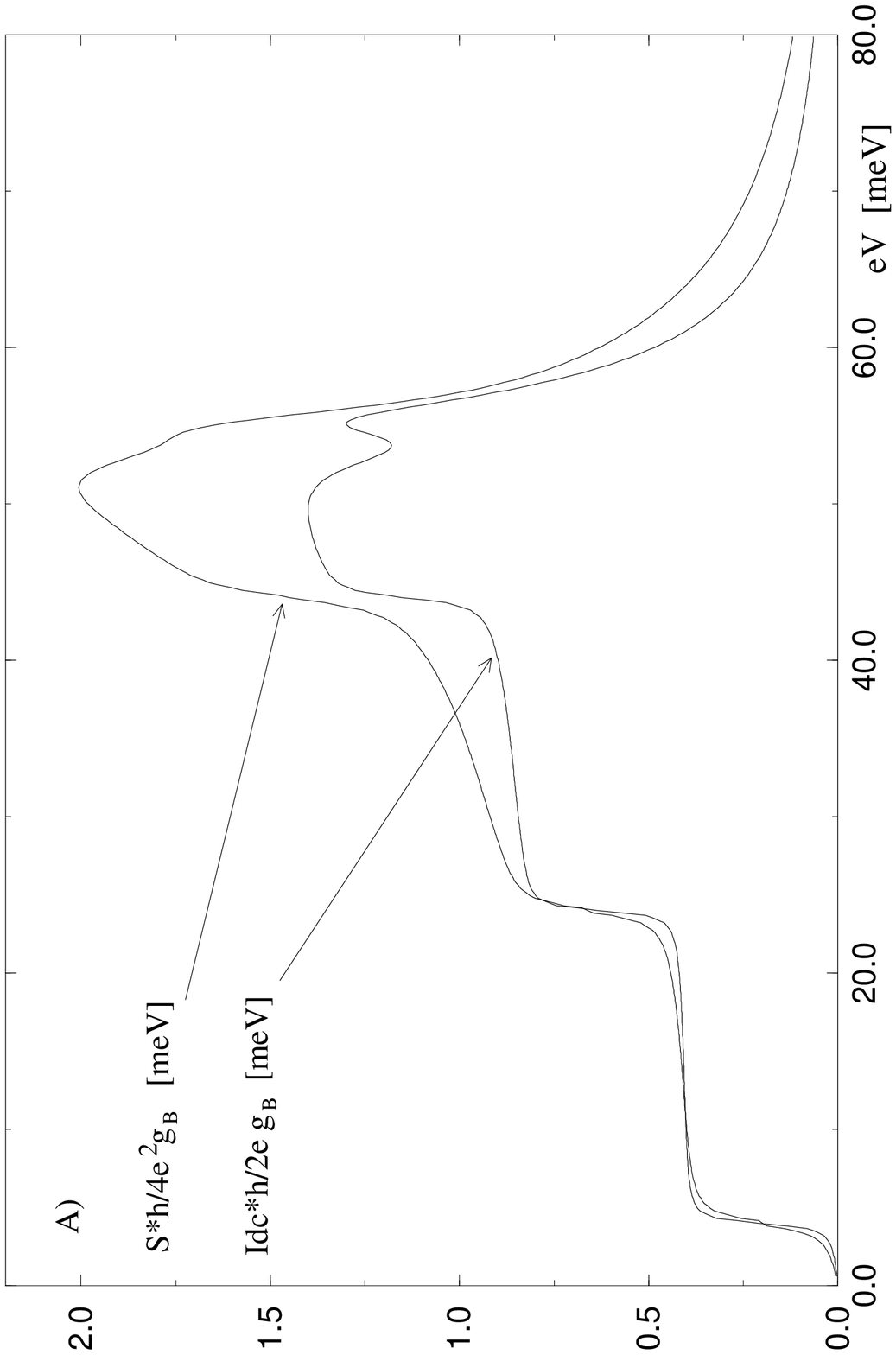,width=7cm}}
\centerline{\psfig{figure=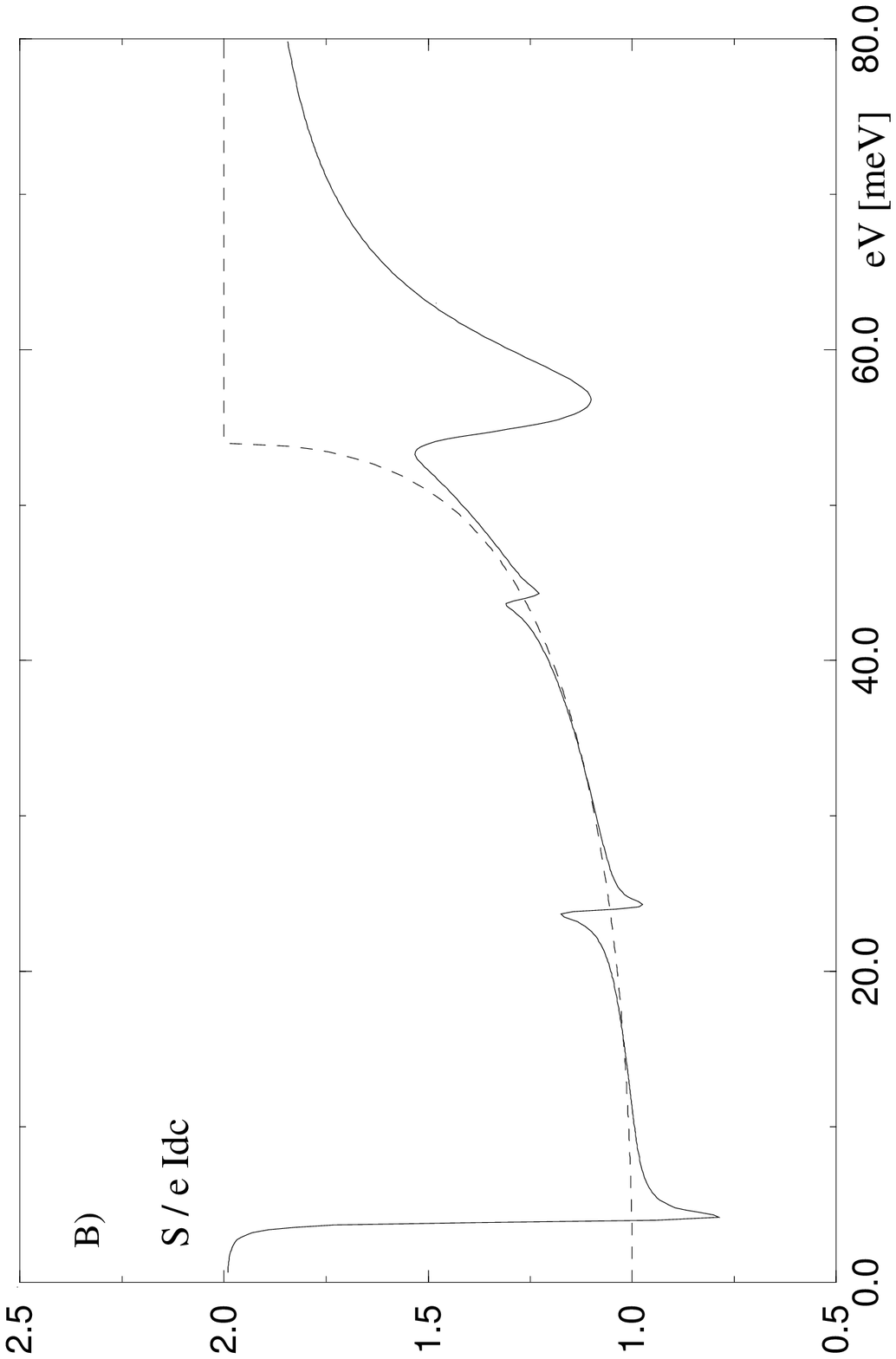,width=7cm}}
\vspace{0.5cm}
\caption{Average current and  zero  frequency  shot  noise (A)  
and dimensionless noise-to-current ratio (B)
for a  symmetric 3D-emitter DBRTS with:
$\Upsilon_e=\Upsilon_c=0.67$  (meV)$^{3/2}$,  
$\hbar\omega_c=10$  meV, $\varepsilon_0=27$  meV, $E_F=30$  meV  
and $\nu=0.5$   meV. 
The shot noise ratio  solid curve was obtained  from the  
exact equation (\protect\ref{s0}), while the dashed curve   
results from the approximation  (\protect\ref{stot1}).}
\par  
\end{figure}  
\newpage
  
\begin{figure}\label{fig4}
\centerline{\psfig{figure=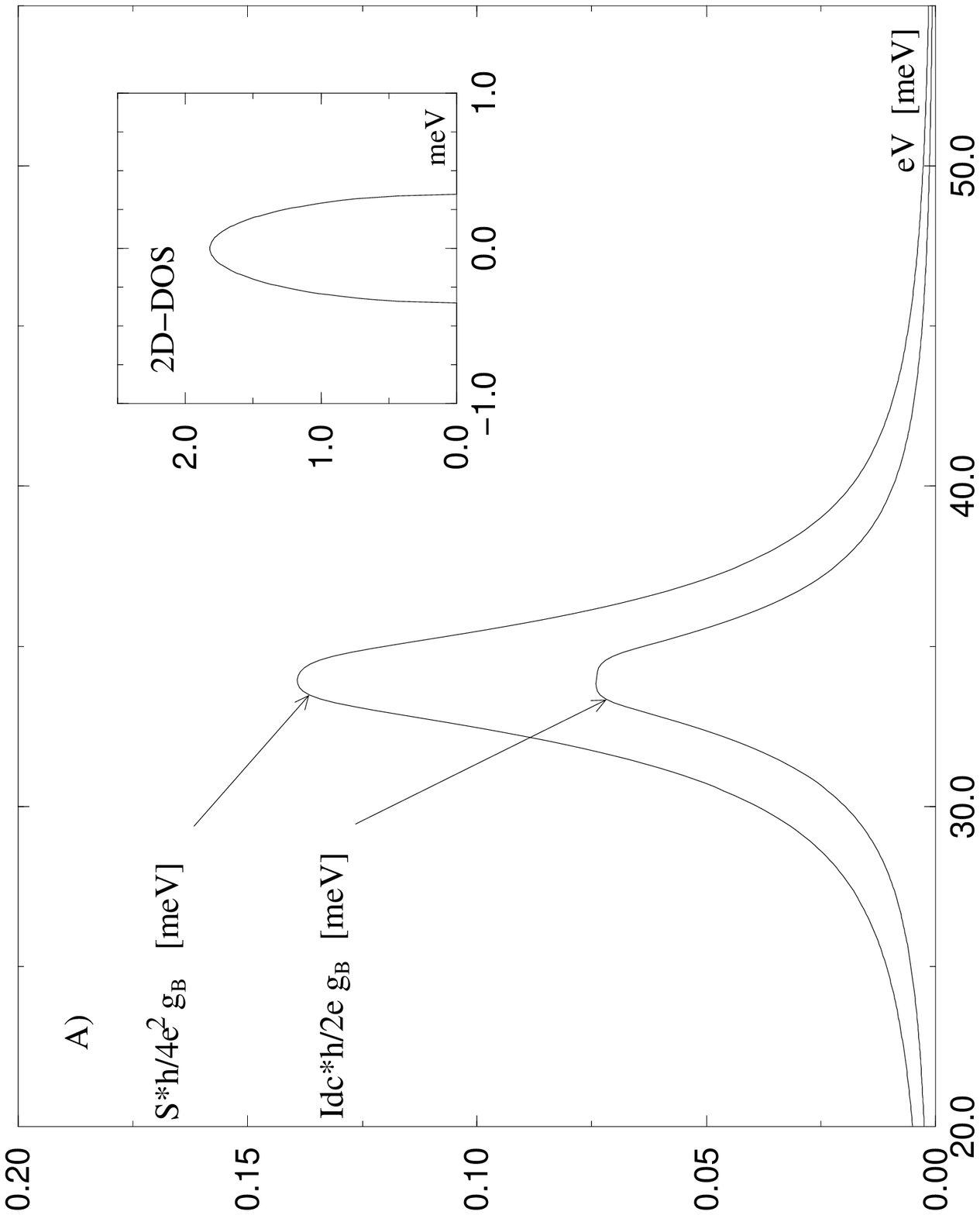,width=8cm}}  
\centerline{\psfig{figure=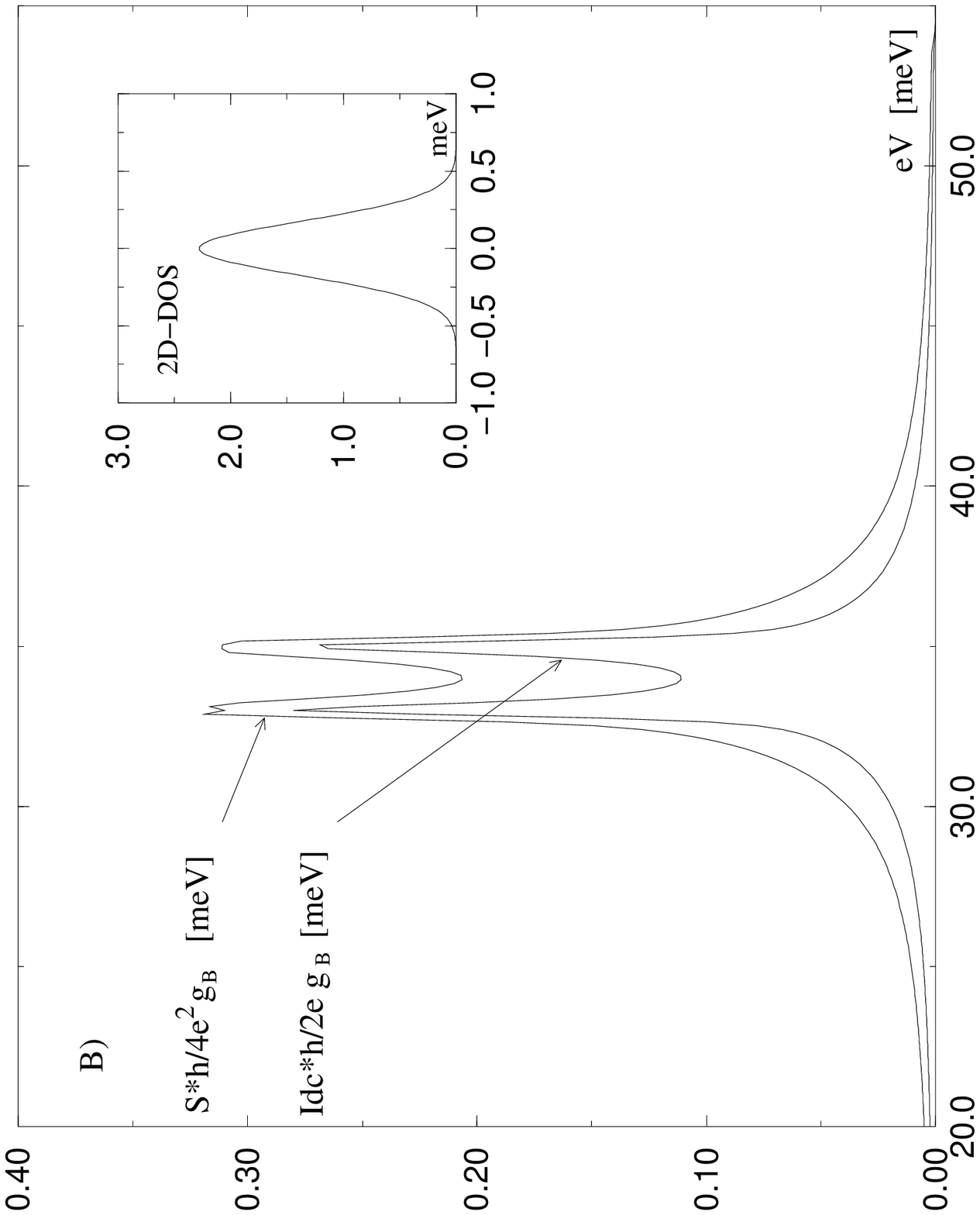,width=8cm}}   
\caption{Average urrent and  zero frequency shot noise  
for a symmetrical 2D-emitter DBRTS with:  
$\Upsilon_e=\Upsilon_c=0.67$ meV$^{3/2}$, $L_{ez}=200\AA$, 
$\hbar\omega_c=10$   meV, $\varepsilon_0 =27$   meV, $E_F=30$   meV,  
$\epsilon_e=10$   meV, $\nu_{3D} =0.5$   meV and $\nu_{2D}=0.35$   meV.  
A) shows the results obtained from a  
semi-elliptic Landau level DOS  
profile in the emitter (see figure inset) while  B) shows the 
corresponding results from a Gaussian DOS profile.}  
\par
\end{figure}
\newpage  
 
\begin{figure}\label{fig5}
\centerline{\psfig{figure=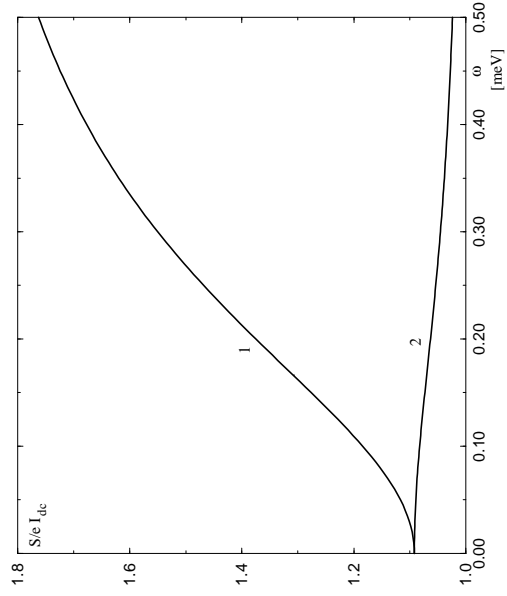,width=8cm}} 
\caption{Frequency dependence of  the noise-to-current ratio  
for a symmetrical DBRTS with the parameters  
$\Upsilon_e=\Upsilon_c=0.67$  meV$^{3/2}$,  
$\hbar\omega_c=10$  meV, $\varepsilon_c=27$  meV, $E_F=30$  meV  
and with the applied voltage $eV=30$ meV.  
Curve `1' shows the case of symmetric barrier capacitances  
(\protect\ref{sw1}), while curve `2' is the result when barrier  
capacitance charge  
accumulation is negligible (\protect\ref{sw2}).}
\par  
\end{figure}
\newpage

\begin{figure}\label{fig6}
\centerline{\psfig{figure=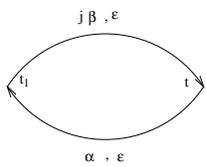,width=3cm}}  
\vspace{0.5cm}
\caption{Diagrammatic representation for the dc-current Green's functions.}
\par  
\end{figure}

\begin{figure}\label{fig7}
\centerline{\psfig{figure=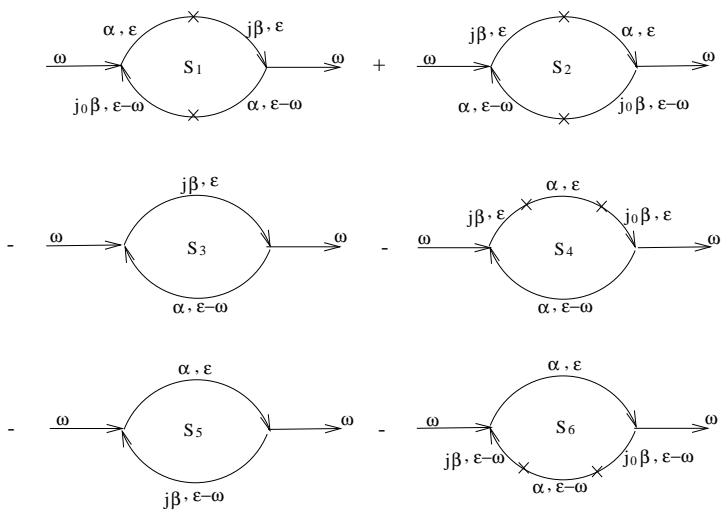,width=8cm}}
\vspace{0.5cm} 
\caption{Diagrammatic representation for the noise Green's functions}
\par  
\end{figure}  
  
\begin{figure}\label{fig8}
\centerline{\psfig{figure=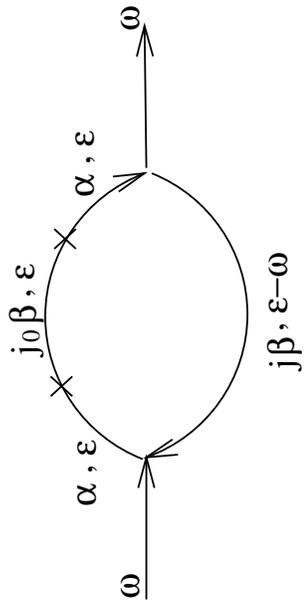,width=4cm}}
\vspace{0.5cm}
\caption{A typical diagram not taken explicitly into account  
since it is already implicitly included in other diagrams.}
\par  
\end{figure}  
\newpage  
  
\end{document}